\documentclass[11pt]{article}

\usepackage{graphicx}
\usepackage{amsmath, amsthm}
\usepackage{amsfonts}
\usepackage{lipsum}
\usepackage{setspace}
\usepackage{natbib}
\setcitestyle{authoryear,open={(},close={)}} 
\usepackage[toc]{appendix}
\usepackage{pgfplots}
\usepackage{tikz}
\pgfplotsset{compat = 1.18}
\usetikzlibrary{positioning, arrows.meta}
\usepgfplotslibrary{fillbetween}
\usepackage{subcaption}
\usepackage{float}

\newtheorem{theorem}{Theorem}
\newtheorem{claim}{Claim}
\newtheorem{lemma}{Lemma}
\newtheorem{proposition}{Proposition}
\newtheorem{corollary}{Corollary}

\renewcommand*{\thefootnote}{\fnsymbol{footnote}}

\title{\bf{ \Large Pricing AI Model Accuracy}}

\author{Nikhil Kumar\footnote{Email: kumarnik@sas.upenn.edu. I am grateful to Rakesh Vohra for his guidance and several helpful discussions. I also thank Krishna Dasaratha, Stefanos Delikouras, Kevin He, participants at the 36th Stony Brook International Conference on Game Theory, and  anonymous reviewers for their helpful comments and suggestions. All remaining errors and omissions are mine. }\\ \textit{University of Pennsylvania}}

\date{November 2025\vspace{-5ex}}

\begin{document}
\fontfamily{cmr}\selectfont
\maketitle	
\vspace{0.2cm}

\normalsize \noindent {\bf Abstract:} This paper examines the market for AI models in which firms compete to provide accurate model predictions and consumers exhibit heterogeneous preferences for model accuracy. We develop a consumer-firm duopoly model to analyze how competition affects firms' incentives to improve model accuracy. Each firm aims to minimize its model's error, but this choice can often be suboptimal. Counterintuitively, we find that in a competitive market, firms that improve overall accuracy do not necessarily improve their profits. Rather, each firm's optimal decision is to invest further on the error dimension where it has a competitive advantage. By decomposing model errors into false positive and false negative rates, firms can reduce errors in each dimension through investments, and they are strictly better off investing on their superior dimension and strictly worse off with investments on their inferior dimension. Profitable investments adversely affect consumers but increase overall market welfare.
\medskip
\renewcommand*{\thefootnote}{\arabic{footnote}}



\doublespacing
\newpage
\section{Introduction}
In the artificial intelligence (AI) and machine learning (ML) industry, an AI model supplier's central goal is to minimize its model's error. This error metric can be very task-dependent; generative AI models prioritize producing outputs strongly resembling the user's intended requests, whereas traditional classification models may try to minimize incorrect predictions on test data. In all such cases, though, firms typically try to minimize some corresponding measure of overall error as a proxy for model performance. 

While this goal is theoretically optimal, achieving a minimal error rate can be both costly and impractical for firms (\citealp{pmlr-v97-zhang19p}). Improving multiple error dimensions simultaneously can be expensive and inefficient. In addition, an improvement in one dimension may lead to a worsening in another dimension, inducing an inverse trade-off across dimensions (\citealp{yaniv1995graininess}). A typical example of this is binary classification, where directly minimizing both Type I and Type II errors with a fixed sample size is typically infeasible (\citealp{mudge2012setting}). Assuming a fixed total level of error, designing a model with fewer false positive (FP) errors would shrink the rejection region but also simultaneously lead to an increase in false negative (FN) error risk, and vice versa. As a result, model designers have to prioritize between FP and FN errors based on their expectations about consumer preferences. A standard statistical approach is fixing one dimension and optimizing the other (\citealp{doi:10.1177/25152459221080396}).

On the demand side, consumers often care less about the total model error and more about the specific types of errors a model makes, as some errors may be more costly to them than others. The bias-variance trade-off is an example of this inverse relation in model performance; lowering bias by improving model complexity typically leads to higher variance (\citealp{geman1992neural}). One specific measure of model performance that consumers often consider is based on FP and FN rates (\citealp{cappelen2023second}). Different consumers exhibit varying levels of sensitivities to FP and FN rates, depending on their specific use cases and their associated cost sensitivities to FP and FN errors.

A prevalent shortcoming with generative AI models is hallucination, where the model generates entirely incorrect outputs. Hallucination rates, or more broadly FP rates, of the most popular models like ChatGPT remain quite significant (\citealp{rawte2023surveyhallucinationlargefoundation}). Users employing AI for creative writing may have a completely different tolerance for hallucination compared to someone using the same model for disease diagnosis. This consumer heterogeneity can be captured by different sensitivities toward hallucination rates. For example, a scientific user may prefer a non-informative model output (i.e., \lq\lq I don't know") as opposed to a model hallucination.

As a motivating example, consider two model users, Company A and Company B, both searching for new employees but struggling to choose new hires. Each company employs an AI-based labor screening model that, for every candidate, outputs either 1 (recommended hire) or 0 (recommended reject). The model therefore has two potential kinds of errors: false positives \textemdash recommending a hire who is actually unsuitable, and false negatives \textemdash rejecting a candidate who would have been a strong hire. Suppose that Company A is a large tech firm, and thus can tolerate a higher volume of applicants; its main concern is missing strong candidates, so false negatives are especially costly. Company B, on the other hand, is a small research lab with very few positions; its main concern is the disruption caused by a bad hire, so false positives are more costly.

This implicit trade-off between FP and FN rates perceived by consumers, along with the difficulties associated with pure error reduction, leads to product differentiation by model suppliers. Such firms specialize through investments towards reducing error rates on certain dimensions to maximize profits; with limited resources, they must choose whether to spend on reducing their FP/FN error rates or expanding their training dataset which reduces the marginal cost of future error improvements. As a result, these supplying firms try to invest in a way that increases profits, attracting more consumers while also maintaining a sufficiently high price. In sum, two relevant phenomena are present in the AI model market: (i) heterogeneous cost sensitivity levels among consumers and (ii) an implicit firm-level trade-off between FP and FN rates. 

In this paper, we develop an economic model that captures both these phenomena and produces a somewhat counter-intuitive result: firms should invest in growing their competitive advantage rather than closing the gap on their inferior error dimension. This result stems from the economic intuition that closing the gap will have a minimal positive effect on market share but a greater negative effect on equilibrium prices, resulting in the firm being worse off. We develop a firm-consumer market model where firms are endowed with an initially fixed error rate, and they can then make investments toward improving either their FP or FN rate. Consumers exhibit heterogeneity through negatively correlated cost sensitivities towards FP and FN rates, which generates a natural trade-off between the two error types.

We define the static version of the model in Section 3, where firms are endowed with fixed FP and FN rates, and equilibrium prices are determined endogenously. Within this fixed error rate framework, we introduce heterogeneity in the cost sensitivities of consumers. Under perfectly positively correlated cost sensitivities, our model simplifies to an extension of the standard Hotelling model. We then consider perfectly negatively correlated sensitivities and analyze the resulting equilibria in the market. Naturally, total firm revenue is maximized when all consumers have negatively correlated sensitivities, as firms are able to differentiate effectively and charge higher prices to attract consumers that match their cost sensitivities. 

Model dynamics are introduced in Section 4, where we consider the stability of the equilibria presented in Section 3. Firms can invest in two ways: directly improving their FP and FN error rates or increasing their firm size, and implicitly the size of their training dataset. We show that the initial market clearing prices are not equilibria; profitable deviations often exist for both firms. The equilibria can be categorized into cases of split domination, where each firm dominates the other on one dimension, versus strict domination, where one firm dominates the other on both dimensions. Under split domination, we present our main result: firms should invest on the error dimension in which they are strictly superior. 

We first consider the case where firms do not face direct costs and show that even in such a simplified case, firms should never invest to close the gap on their inferior dimension. We then introduce an AI-inspired cost function and show that the optimal investment decisions can be explicitly tied to the firm's scalability and growth potential. We also establish that such profitable investments for firms adversely affect overall consumer welfare, as increased prices result in a higher average cost per consumer.

\section{Related Literature}

Our work builds upon past research at the intersection of machine learning and strategic behavior, focusing on how competition affects efficiency and the resulting equilibria. Previous literature has extensively analyzed competition in such settings. For example, \citet{10.1145/1993636.1993666} propose an optimization-based framework for strategic competition, but does not include prices. \citet{feng2022bias} examine the bias-variance trade-off in competitive environments, demonstrating a tendency for firms to favor variance-based errors over bias-based errors. Similarly, \citet{jagadeesan2024improvedbayesriskyield} propose a classification model under which competition can potentially lead to worse overall model accuracy. Other studies adopt more information-driven approaches to competition (\citealp{liang2020gamesincompleteinformationplayed}) and explore coordination among firms to learn about consumer types (\citealp{gradwohl2023coopetition}).

We contribute to the long-standing industrial organization literature by presenting a model of horizontal differentiation in an AI model market. Consumers' cost sensitivities govern their purchasing behavior of models and thus determine market equilibria. \cite{shaked1982relaxing} present a multi-stage game model with vertical differentiation, but assume consumers have identical tastes. \cite{gabszewicz1979price} also assume no horizontal differentiation but discuss the notion of firm specialization. 

This paper differs from the related literature along multiple dimensions. First, we focus explicitly on competition over false positive (FP) and false negative (FN) rates rather than on the bias-variance trade-off. While the bias-variance framework is instrumental for understanding model complexity and generalization (\citealp{feng2022bias}), the FP/FN trade-off is more specific to classification tasks. \citet{jagadeesan2024improvedbayesriskyield} examines competition based on accuracy measured with Bayes risk, but firms do not choose prices. Classification problems are central to many real-world applications, such as medical diagnostics, fraud detection, and content filtering, where FP and FN errors carry distinct and significant costs. By concentrating on FP and FN rates, we enable a more granular analysis of error distributions and consumer preferences. 

Unlike \citet{feng2022bias}, where firms optimize directly over bias and variance-based errors, the FP and FN rates in our model simultaneously determine both the first and second moments of the cost function. Consumers care about the expected cost of a mistake, and a lower expected cost does not necessarily imply a higher variance. This approach is particularly effective for modeling consumer behavior, making it easier to see how different types of mistakes affect consumer satisfaction. 

Second, we extend the existing literature by incorporating consumer market share and pricing dynamics into the model. By introducing a consumer distribution and corresponding market share, we allow firms to compete not only by optimizing their error rates but also by strategically setting prices to attract consumers. In our model, consumers choose between firms based on their individual sensitivities to FP and FN errors as well as the prices charged by each firm. This resulting structure relies on market clearing conditions and endogenous price equilibria. Firms must balance the trade-off between improving model accuracy (reducing FP and FN rates) and setting competitive prices to capture a larger market share. By explicitly modeling these interactions, we link firms' algorithmic decisions under competition to pricing and capturing market share. 

Third, we introduce heterogeneity among both consumers and firms. Empirical evidence suggests that consumers exhibit varying sensitivities to FP and FN errors (\citealp{cappelen2023second}). We formally model this heterogeneity using FP and FN cost parameters, accounting for cardinal differences, the magnitude of costs associated with FP and FN errors for each consumer, and ordinal differences, the relative importance or priority that consumers assign to FP versus FN errors. Similarly, we incorporate firm heterogeneity through varying error endowments. Based on their initial endowments, firms may specialize in optimizing for different error types or target specific consumer segments, leading to potentially diverse equilibria.

\section{The Model}

Consumers are uniquely defined by a pair of cost sensitivity parameters $\alpha$ and $\beta$, both $\sim U[0, 1]$. $\alpha $ captures the agent's cost sensitivity to false positives, whereas $\beta$ captures cost sensitivity to false negatives. These parameters are private information; only consumers know their own cost sensitivities. Consumer heterogeneity is thus modeled through variation in $\alpha$ and $\beta$.

Firms are endowed with a false positive (FP) rate and a false negative (FN) rate. We assume that $FP, FN \geq 0$ for all firms and assume that there is product differentiation, so that no two firms have the same pair of FP and FN rates. A firm's FP and FN rates are public information to consumers and other firms. We also assume that firms' FN rates are less volatile than FP rates, meaning differences in FN rates across firms are smaller than differences in FP rates: $|FN_2 - FN_1| < |FP_2 - FP_1|$. This assumption imposes an ordering on the volatility of FP and FN rates; however, the model and presented results all hold under the reverse ordering as well. 

Each firm $j$ chooses a non-negative price $p_j \in [0, \Bar{p}]$ to maximize revenues, and market share to each firm is calculated as a proportion of the unit interval $[0, 1]$. For simplicity, we assume a duopoly model structure with two competing firms in the market. We assume firms do not incur any direct costs from purely producing their model; however, they face direct investment and development costs dependent on their FP and FN rates as well as a underlying firm size parameter. We discuss this more in detail in Section \ref{section:flex}. 

Consumers buy models from firms at market prices, and their primary objective is to minimize total cost. Consumer $ i$'s cost function $c_i$ when buying from firm $j$ is the corresponding weighted error sum plus price, defined as follows:
\begin{equation}
\begin{aligned}
    c_i(j) = (\alpha_i \times FP_j) + (\beta_i \times FN_j) + p_j,
\end{aligned}
\end{equation}
where the $i$ subscripts denote the consumer's personalized sensitivities for FPs and FNs, respectively, and the $j$ subscripts denote the firm's error rates and chosen prices. Each consumer purchases from the firm that minimizes cost: 
\[
\begin{aligned}
& \underset{j}{\text{min}}
& & c_i(j) \hspace{0.2cm}\Rightarrow & \underset{j}{\text{min}}
& & (\alpha_i \times FP_j) + (\beta_i \times FN_j) + p_j.
\end{aligned}
\]
Market clearing conditions and equilibrium prices are determined correspondingly. 

\subsection{Perfectly Positively Correlated Cost Sensitivities}
We now impose structure on consumers' cost sensitivities, first considering the case where $\alpha = \beta$, implying perfectly positively correlated consumer sensitivities to FP and FN rates. Under this structure, the model simplifies to the Hotelling model (\citealp{hotelling1981stability}; \citealp{d1979hotelling}). For conciseness, we define $\mathcal{F}_j$ as the total error rates for firm $j$: $\mathcal{F}_j = FP_j + FN_j$. The consumer's cost function $c_i(j)$ then simplifies to:
\begin{equation}
\begin{aligned}
    c_i(j) &= \alpha_iFP_j + \alpha_i FN_j + p_j\\
    & = \alpha_i\mathcal{F}_j + p_j. \nonumber
\end{aligned}
\end{equation}

\begin{claim}[\text{Price Equilibrium with $\alpha = \beta$}]
Equilibrium prices under perfectly positively correlated cost sensitivities are dependent only on firms' total error rates.
\[
\begin{array}{c@{\hskip 2cm}c@{\hskip 2cm}c}
    \textit{If } \mathcal{F}_1 - \mathcal{F}_2 > 0: & \textit{If } \mathcal{F}_1 - \mathcal{F}_2 < 0: & \textit{If } \mathcal{F}_1 - \mathcal{F}_2 = 0: \\
    p_1 = \frac{1}{3} (\mathcal{F}_1 - \mathcal{F}_2) & p_1 = \frac{2}{3} (\mathcal{F}_2 - \mathcal{F}_1) & p_1 = \Bar{p} \\
    p_2 = \frac{2}{3} (\mathcal{F}_1 - \mathcal{F}_2) & p_2 = \frac{1}{3} (\mathcal{F}_2 - \mathcal{F}_1) & p_2 = \Bar{p}
\end{array}
\]
\end{claim}

The resulting equilibrium prices depend solely on each firm's total error rates $\mathcal{F}_j$. The firm with lower overall error rates charges a higher price in the market equilibrium as its better product appeals to a larger market share. In particular, the firm with a lower total error chooses a price twice as high as the other firm. When both firms have the same total error rates, they charge the same price and choose the highest possible price they can charge: $\Bar{p}$. Precise derivations are provided in Appendix A.1. 

\subsection{Perfectly Negatively Correlated Cost Sensitivities} 

We now consider perfectly negatively correlated cost sensitivities by letting $\alpha = 1 - \beta$. This specification implies that a consumer's sensitivities for FP and FN rates are negatively linear in one another. Similar to Section 3.1, we define $\Delta F_j$ as the difference between a firm's FP and FN rate for conciseness: $\Delta F_j = FP_j - FN_j$. Consumer $i$'s cost function $c_i(j)$ can be expressed as:
\begin{equation}
\begin{aligned}
    c_i(j) &= (\alpha_i \times FP_j) + ((1-\alpha_i) \times FN_j) + p_j \\
    &= \alpha_i\Delta F_j + p_j + FN_j \nonumber
\end{aligned}
\end{equation}

To proceed, we specify an ordering on the FP and FN rates of both firms. We consider two cases: strict domination and split domination. Under split domination, each firm dominates the other on one dimension, leading to non-zero prices for both firms. Under strict domination, one firm has strictly lower FP and FN rates than the other and thus effectively dominates the other firm. We show that the inferior firm always exits the market, and the dominating firm operates as a monopoly. Note that ordering on FP and FN rates does not affect the model in Section 3.1, as no distinction exists between FP and FN rates and, therefore, only total error is relevant.

\subsubsection{Split Domination}

We assume a strict ordering on the FP and FN rates of both firms. In particular, we impose that firm 1 has a lower FN rate and higher FP rate than firm 2: $FN_1 < FN_2 \hspace{0.1cm}\text{and} \hspace{0.1 cm} FP_1 > FP_2$. In practical terms, this can be thought of as firm 1 specializing in lowering its FN errors while firm 2 lowers its FP errors. Note that a reverse ordering on error rates leads to the same results due to symmetry.

The resulting firm reaction functions are as follows: 
\begin{equation}
    \begin{aligned}
    p_1^*(p_2) = \max \{0, \frac{FN_2 - FN_1 + p_2}{2}\} \\
    p_2^*(p_1) = \max\{0, \frac{FP_1 - FP_2 + p_1}{2}\} \nonumber
    \end{aligned}
\end{equation}
The reaction functions above indicate a level of asymmetry, as firm 1's choice is dependent only on the difference in FN rates across firms, whereas firm 2's choice is dependent on the difference in FP rates. This asymmetry in reaction functions is implicitly due to the definition of $\alpha$ and the consequent indifferent consumer $\Tilde{\alpha}$. In particular, a larger $\Tilde{\alpha}$ implies that firm 1 gets a larger market share, and a lower $\Tilde{\alpha}$ gives firm 2 a larger market share. Thus, reaction functions stem from the concrete structure of $\Tilde{\alpha}$, which is increasing in $FN_2 - FN_1$ and decreasing in $FP_1 - FP_2$. Derivations and analysis are presented in Appendix A.2. 

\begin{claim}[\text{Price Equilibrium with $\alpha = 1- \beta$ and Split Domination}]

Equilibrium prices under perfectly negatively correlated cost sensitivities are dependent solely on ordinal differences in FP and FN rates:
\begin{align*}
    p_1^* &= \frac{2(FN_2 - FN_1) + FP_1 - FP_2}{3} \\
    p_2^* &= \frac{FN_2 - FN_1 + 2(FP_1 - FP_2)}{3}
\end{align*}
\end{claim}

We observe that equilibrium prices are unaffected by the numerical value of the FP and FN rates but rather by the competitive advantage one firm has over the other. Note that prices are always strictly positive due to the imposed ordering on FP and FN rates above; $p_1^*, p_2^*$ are strictly positive because the differences $FN_2 - FN_1$ and $FP_1 - FP_2$ are always positive.

Figure 1 plots the reaction functions for both firms under certain fixed parameter values of FP and FN rates. The pair of price equilibria is always non-negative under the assumptions above; we show that if the assumptions are violated, one firm sets prices to 0, and the other operates as a monopoly. 

\begin{figure}[H] 
    \centering
    \begin{subfigure}[b]{0.45\textwidth}
        \begin{tikzpicture}
        \begin{axis}[
            scale = 0.7,
            xmin = 0, xmax = 1,
            ymin = 0, ymax = 1,
            axis lines = left,
            xlabel={$p_1$},
            ylabel={$p_2$},
            legend pos=north west,
            legend style = {font=\small, fill opacity = 0.9, rounded corners = 4 pt}
        ]
        \addplot[color = red, thick, domain=0:1] {2*x - 0.1};
        \addlegendentry{$p_1 = 0.5p_2 + 0.05$}
        
        \addplot[color = blue, thick, domain=0:1] {0.5*x + 0.05};
        \addlegendentry{$p_2 = 0.5p_1 + 0.05$}
        
        \node at (axis cs:0.1, 0.1) [circle,fill=black,inner sep=1.5pt,label={[shift={(0.25,-0.2)}]90:{(0.1, 0.1)}}] {};
        \end{axis}
        \end{tikzpicture}
        \caption{$FP_1 = FN_1 = 0.2, FP_2 = 0.1, FN_2 = 0.3$; Symmetric non-negative  equilibrium prices}
    \end{subfigure}
    \hfill
    \begin{subfigure}[b]{0.45\textwidth} 
        \begin{tikzpicture}
        \begin{axis}[
            scale = 0.7,
            xmin = 0, xmax = 1,
            ymin = 0, ymax = 1,
            axis lines = left,
            xlabel={$p_1$},
            ylabel={$p_2$},
            legend pos=north west,
            legend style = {font=\small, fill opacity = 0.9, rounded corners = 4 pt}
        ]
        \addplot[color = red, thick, domain=0:1] {2*x - 0.05};
        \addlegendentry{$p_1 = 0.5p_2 + 0.025$}
        
        \addplot[color = blue, thick, domain=0:1] {0.5*x + 0.05};
        \addlegendentry{$p_2 = 0.5p_1 + 0.05$}

        \node at (axis cs:0.2/3, 0.25/3) [circle,fill=black,inner sep=1.5pt,label={[shift={(0.25,-0.15)}]90:{(0.07, 0.08)}}] {};
        
        \end{axis}
        \end{tikzpicture}
        \caption{$FP_1=0.3, FP_2 = 0.2, FN_1 = 0.05, FN_2 =0.1$; Asymmetric non-negative equilibrium prices}
    \end{subfigure}
    \begin{subfigure}[b]{0.45\textwidth} 
        \vspace{0.2cm}
        \begin{tikzpicture}
        \begin{axis}[
            scale = 0.7,
            xmin = 0, xmax = 1,
            ymin = 0, ymax = 1,
            axis lines = left,
            xlabel={$p_1$},
            ylabel={$p_2$},
            legend pos=north west,
            legend style = {font=\small, fill opacity = 0.9, rounded corners = 4 pt}
        ]
        \addplot[color = red, thick, domain=0:1] {2*x + 0.1};
        \addlegendentry{$p_1 = 0.5p_2 - 0.05$}
        
        \addplot[color = blue, thick, domain=0:1] {0.5*x - 0.05};
        \addlegendentry{$p_2 = 0.5p_1 - 0.05$}

        \end{axis}
        \end{tikzpicture}
        \caption{$FP_1 = FN_1 = 0.2, FP_2 = 0.1, FN_2 = 0.3$; no equilibrium due to violated assumption}
    \end{subfigure}
    \hfill
    \begin{subfigure}[b]{0.45\textwidth} 
        \begin{tikzpicture}
        \begin{axis}[
            scale = 0.7,
            xmin = 0, xmax = 1,
            ymin = 0, ymax = 1,
            axis lines = left,
            xlabel={$p_1$},
            ylabel={$p_2$},
            legend pos=north west,
            legend style = {font=\small, fill opacity = 0.9, rounded corners = 4 pt}
        ]
        \addplot[color = red, thick, domain=0:1] {2*x - 0.01};
        \addlegendentry{$p_1 = 0.5p_2 + 0.005$}
        
        \addplot[color = blue, thick, domain=0:1] {0.5*x + 0.25};
        \addlegendentry{$p_2 = 0.5p_1 + 0.25$}
        
        \node at (axis cs:0.52/3, 1.01/3) [circle,fill=black,inner sep=1.5pt,label={[shift={(0.13,-0.3)}]90:{(0.17, 0.34)}}] {};
        \end{axis}
        \end{tikzpicture}
        \caption{$FP_1 = 0.6, FP_2 = 0.1, FN_1 = 0.01, FN_2 = 0.02$; equilibrium with large price disparity}
    \end{subfigure}
    \vspace{0.2cm}
    \caption{Price equilibria under split domination with varying fixed FP and FN rate values.}
\end{figure}
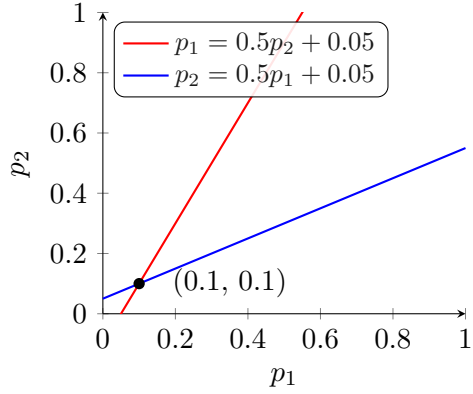
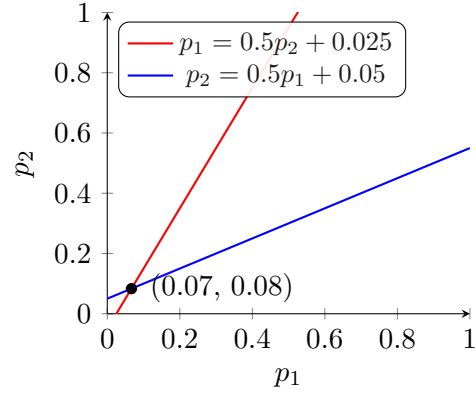
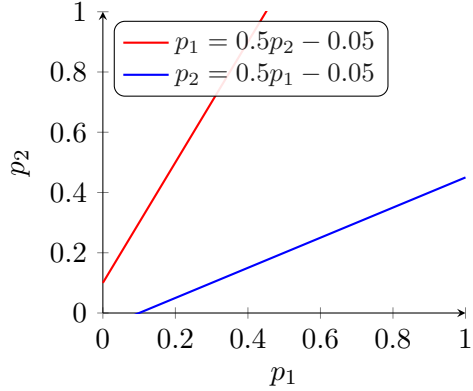
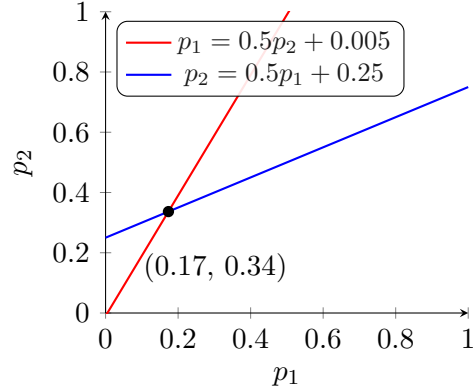

\subsubsection{Strict Domination}

We assume that firm 1 strictly dominates firm 2, meaning that $FP_1 < FP_2$ and $FN_1 < FN_2$. Firm 1 thus has strictly lower errors on both the FP and FN dimensions compared to firm 2, meaning its product is strictly better than firm 2. Note that a symmetrical analysis can be conducted with firm 2 dominating firm 1.

Using the assumption that FN rates are less volatile than FP rates: i.e. $|FN_2 - FN_1| < |FP_2 - FP_1|$, firm 2 exits the market, and firm 1 operates as a monopoly. 

\begin{claim}[\text{Monopoly Under Strict Domination}]
    Assume $|FN_2 - FN_1| < |FP_2 - FP_1|$. Then, firm 2 exits the market, and firm 1 sets prices proportional to the difference in FN rates. 
\end{claim}
In particular, $p_1^* = \frac{FN_2 - FN_1}{2}$ and $p_2^* = 0$. As the difference in FN rates between firm 1 and firm 2 becomes smaller, firm 1 charges a lower and lower price in equilibrium to keep firm 2 out of the market. Thus, the equilibrium price $p_1$ is increasing in the FP and FN rate differences between firms. Choosing a higher $p_1$ would give firm 2 a non-zero market share, which would lower firm 1's revenue. The reaction functions are identical to the split domination case above, but the lower bound of 0 becomes more prevalent. More details are provided in Appendix A.3, and we present a couple of examples in Figure 2. 

\begin{figure}[H] 
    \centering
    \begin{subfigure}[b]{0.45\textwidth}
        \begin{tikzpicture}
        \begin{axis}[
            scale = 0.7,
            xmin = 0, xmax = 1,
            ymin = 0, ymax = 1,
            axis lines = left,
            xlabel={$p_1$},
            ylabel={$p_2$},
            legend pos=north west,
            legend style = {font=\small, fill opacity = 0.9, rounded corners = 4 pt}
        ]
        \addplot[color = red, thick] {2*x - 0.05};
        \addlegendentry{$p_1 = \max(0, 0.025 + 0.5p_2)$}
        
        \addplot[color = blue, thick, domain=0:0.1] {0};
        \addplot[color = blue, thick, domain=0.1:1] {-0.05 + 0.5*x};
        \addlegendentry{$p_2 = \max(0, -0.05 + 0.5p_1)$}

        \node at (axis cs:0.025, 0) [circle,fill=black,inner sep=1.5pt,label={[shift={(0.3,-0.2)}]90:{(0.1, 0.1)}}] {};
        
        \end{axis}
        \end{tikzpicture}
        \caption{$FP_1 = 0.1, FP_2 = 0.2, FN_1 = 0.05, FN_2 = 0.1$; Monopoly with small $p_1^* = 0.025$.}
    \end{subfigure}
    \hfill
    \begin{subfigure}[b]{0.45\textwidth} 
        \begin{tikzpicture}
        \begin{axis}[
            scale = 0.7,
            xmin = 0, xmax = 1,
            ymin = 0, ymax = 1,
            axis lines = left,
            xlabel={$p_1$},
            ylabel={$p_2$},
            legend pos=north west,
            legend style = {font=\small, fill opacity = 0.9, rounded corners = 4 pt}
        ]
        \addplot[color = red, thick] {2*x - 0.3};
        \addlegendentry{$p_1 = \max(0, 0.15 + 0.5p_2)$}
        
        \addplot[color = blue, thick, domain=0:0.4] {0};
        \addplot[color = blue, thick, domain=0.4:1] {-0.2 + 0.5*x};
        \addlegendentry{$p_2 = \max(0, -0.2 + 0.5p_1)$}

        \node at (axis cs:0.15, 0) [circle,fill=black,inner sep=1.5pt,label={[shift={(0.3,-0.2)}]90:{(0.15, 0)}}] {};
        
        \end{axis}
        \end{tikzpicture}
        \caption{$FP_1 = 0.1, FN_1 = 0.05, FP_2 = 0.5, FN_2 = 0.35$; Monopoly with larger $p_1^* = 0.15$.}
    \end{subfigure}
    \vspace{0.2cm}
    \caption{Price equilibria under strict domination with varying fixed FP and FN rate values.}
\end{figure}
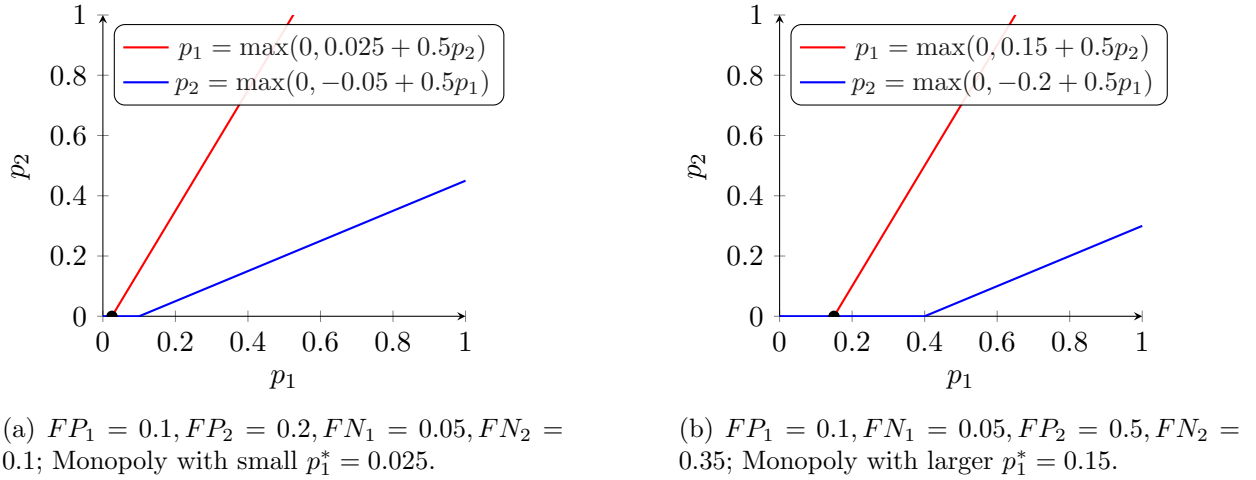

\subsection{Combination of Positively and Negatively Correlated Sensitivities}

We now consider a combination of the perfectly positively correlated model and perfectly negatively correlated model through a proportion parameter $\zeta$. We assume that all consumers have either $\alpha = \beta$ or $\beta = 1 - \alpha$, and $\zeta \in [0, 1]$ defines the proportion of the overall consumer population with $\alpha = \beta$. We assume the split domination case. 

\begin{claim}[\text{Relationship Between Prices and Consumer Heterogeneity}]
    $p_1^*$ and $p_2^*$ are strictly concave in $\zeta$. 
\end{claim}

This trivially implies that prices and revenue are highest when consumers all have perfectly negatively correlated cost sensitivities, i.e., when $\zeta = 0$. See  Appendix \ref{appendix:graphs} for the relevant graph of firm prices and revenues across varying $\zeta$. This result stems from the fact that when $\alpha = \beta$, consumers exhibit no preferential distinction between the FP and FN rates themselves. The market share for each firm remains significant even when prices are pushed up, so revenue increases. 

When differentiation exists ($\alpha = 1- \beta$), firms are able to charge a higher price and still attract a non-trivial proportion of the market share, as certain consumers will buy almost exclusively from a certain firm due to their type. In particular, consider firm 1, which specializes in FN rates. Consumers who have very high-cost sensitivities for FN rates (high $\beta$) have a higher willingness to pay to avoid buying from firm 2. Therefore, they will buy from firm 1 even under higher prices. 

\section{Firm Flexibility in FP and FN Rates}
\label{section:flex}

In Section 3, we assume that firms are endowed with fixed FP and FN rates and have no ability to change them. We now introduce flexibility on the error dimension through investment; firms can invest to reduce their FP and FN rates. We consider such investments for both firms under both the strict and split domination cases and analyze the existence and direction of profitable deviations. 

\subsection{Investments for Inferior Firm in Strict Domination}

As mentioned in Claim 3, the strictly dominating firm (firm 1) operates as a monopoly under the strict domination model specification. Therefore, investments from firm 1 towards lower error rates will have no impact on revenues. Firm 2 will be even less incentivized to enter the market, and firm 1 already captures the entire market share, so its revenue is unchanged in FP and FN rates: $\frac{\partial R_1}{\partial FN_1} = \frac{\partial R_1}{\partial FP_1} = 0$. 

Since firm 2 initially does not enter the market, they have zero revenue. Therefore, an investment that captures a strictly positive proportion of the market share is profitable for firm 2. We show that firm 2 can make strictly positive revenue by investing on its FP and FN rates; however, the sizes of such beneficial investments vary by dimension and error rate differences. The distinction between investments in FP and FN rates stems from the volatility of error rates.

\begin{proposition}[\text{Inferior Firm Investments on FN Rates}]
    Assume $|FN_2 - FN_1| < |FP_2 - FP_1|$, and firm 1 strictly dominates firm 2 on both dimensions. An investment by firm 2 which lowers $FN_2$ to $FN_2$\textsuperscript{\textdagger} results in strictly positive revenue iff the investment achieves FN domination: i.e. $FN_2$\textsuperscript{\textdagger} $ < FN_1$.
\end{proposition}

Intuitively, if an investment achieves domination on the FN dimension, we transition from strict domination to split domination. Then, by Claim 2, both firms must charge strictly positive prices, which then implies a non-negative market share for both firms. Similarly, if an investment results in strictly positive revenue for firm 2, it implies strictly positive chosen prices and market share. For both such conditions to hold, firm 2 must thus improve its FN rate to be better than firm 1. 

We now show that investments in the FP dimension have potentially greater implications. 

\begin{theorem}[\text{Inferior Firm Investments on FP Rates}]
    Assume $|FN_2 - FN_1| < |FP_2 - FP_1|$, and firm 1 strictly dominates firm 2 on both dimensions. If an investment by firm 2 lowers $FP_2$ to $FP_2$\textsuperscript{\textdagger} such that $|FP_2\textsuperscript{\textdagger} - FP_1 | < \frac{1}{2}|FN_2-FN_1|$, then the investment results in strictly positive revenue.
\end{theorem}

This relationship can be intuitively understood in Figure 3; a sufficient decrease in $FP_2$ shifts firm 2's reaction function to the left such that the intersection point has both non-zero elements. The resulting equilibrium makes firm 2 strictly better off and firm 1 strictly worse off (the negative effect in demand is stronger than the positive effect on prices). We present further details of the theorem and analysis in Appendix B.1. 

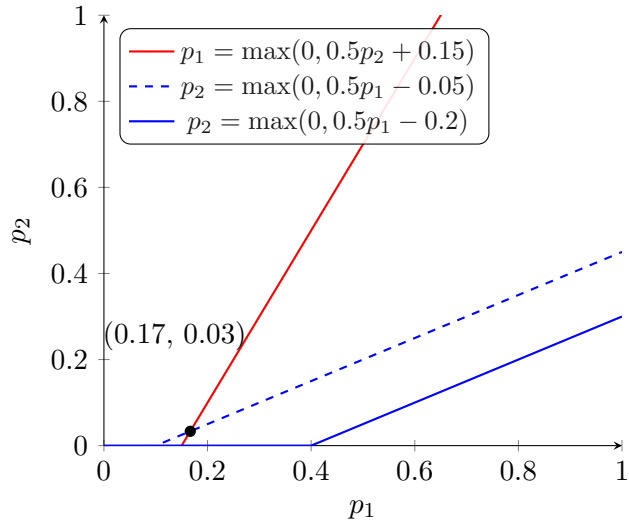
\begin{figure}[H] 
    \centering
    \begin{tikzpicture}
    \begin{axis}[
        scale = 1,
        xmin = 0, xmax = 1,
        ymin = 0, ymax = 1,
        axis lines = left,
        xlabel={$p_1$},
        ylabel={$p_2$},
        legend pos=north west,
        legend style = {font=\small, fill opacity = 0.9, rounded corners = 4 pt}
    ]

    \addplot[color = red, thick, domain=0:1] {2*x - 0.3};
    \addlegendentry{$p_1 = \max(0, 0.5p_2 + 0.15)$}

    \addplot[color = blue, thick, dashed, domain=0:1] {0.5*x - 0.05};
    \addlegendentry{$p_2 = \max(0, 0.5p_1 - 0.05)$}

    \addplot[color = blue, thick, domain=0:0.4] {0};
    \addplot[color = blue, thick, domain=0.4:1] {-0.2 + 0.5*x};
    \addlegendentry{$p_2 = \max(0, 0.5p_1 - 0.2)$}

   \node at (axis cs:0.1667, 0.0333) [circle,fill=black,inner sep=1.5pt,label={[shift={(-0.03,-0.15)}]90:{(0.17, 0.03)}}] {};
   
    \end{axis}
    \end{tikzpicture}
    \caption{Effect on equilibrium prices when firm 2 reduces $FP_2$ from 0.5 to 0.2: $(p_1^*, p_2^*) = (0.15, 0) \Rightarrow (0.167, 0.033)$.}
\end{figure}

Although investments in both dimensions are potentially beneficial, the optimal investment from the inferior firm's point of view is dependent on the distance it has to travel in both dimensions to obtain strictly positive revenue. 

\begin{proposition}[\text{Optimal Investment Dimension}]
    If $\frac{3}{2}|FN_1 - FN_2| > |FP_1 - FP_2|$, then the inferior firm should invest on the FP dimension. Similarly, if $\frac{3}{2}|FN_1 - FN_2| < |FP_1 - FP_2|$, then the inferior firm should invest on the FN dimension. If the two are equal, the firm is indifferent.
\end{proposition} 

The intuition for this proposition can be clearly seen in Figure 4, where the relevant arrows signify the required change in the error rate differences to have strictly positive revenue. To determine the optimal dimension to invest in, the inferior firm must compare the amount it must invest in either dimension to get strictly positive profits. Using Proposition 1 and Theorem 1, investment costs for the FP and FN dimension are depicted in red and blue, respectively; the optimal choice is dependent on which investment cost is larger. 

Note that the investment with a higher return can be determined by the initial differences in FP and FN rates. Thus, if the volatility of FN rates is significantly smaller than FP rates, then reducing the FN rate would require smaller investments than a corresponding increase in FP rates. Similarly, if FN rates are almost as volatile as FP rates, then investing on the FP dimension would be essentially half as expensive as an FN rate investment.  
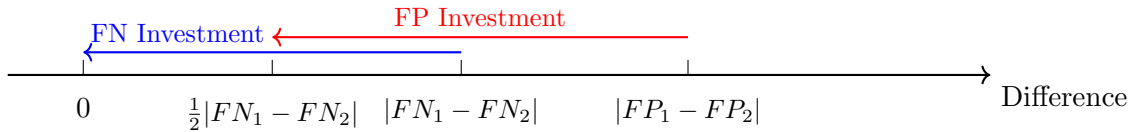
\begin{figure}[h!]
    \centering
    \begin{tikzpicture}
        \draw[thick, ->] (-1,0) -- (12,0) node[anchor=north west] {Difference};
        \draw (0,0) -- (0,0.2);
        \node[below=5pt] at (0, 0) {$0$};
    
        \draw (2.5,0) -- (2.5,0.2);
        \node[below=5pt] at (2.5, 0) {\small $\frac{1}{2}|FN_1 - FN_2|$};
        
        \draw (5,0) -- (5,0.2);
        \node[below=5pt] at (5, 0) {\small $|FN_1 - FN_2|$};
    
        \draw (8,0) -- (8,0.2);
        \node[below=5pt] at (8, 0) {\small $|FP_1 - FP_2|$};
        
        \draw[->, thick, red] (8, 0.5) -- (2.5, 0.5) node[midway, above] {\small FP Investment};
        \draw[->, thick, blue] (5, 0.3) -- (0, 0.3) node[pos=0.75, above] {\small FN Investment};
    \end{tikzpicture}
    \caption{Investment distances to reach strictly positive profits.}
    \label{fig:investment-directions}
\end{figure}

\subsection{Investments Under Split Domination}

We now consider investments in the split domination case, where neither firm strictly dominates the other. As in Section 3.2, assume that $FN_1 <  FN_2, FP_1 > FP_2$: firm 1 dominates in FN rates, and firm 2 dominates in FP rates. Each firm has two potential profitable deviations: invest on the strictly dominated dimension or the strictly dominating dimension. More concretely, firm 1 may try to invest to strengthen its advantage over firm 2 in FN rates, or it may try to close the gap to firm 2 on FP rates. A similar pair of choices exist for firm 2. 

\begin{lemma}[\text{Investments in Dominated Dimension}]
    Firm 1's revenue is strictly increasing in FP rates, and firm 2's revenue is strictly increasing in FN rates. This implies that investing on the strictly dominated dimension results in a strict decrease in revenue. 
\end{lemma}

This lemma implies that $\frac{\partial R_1}{\partial FP_1} > 0$ and $\frac{\partial R_2}{\partial FN_2} > 0$. Since firms invest to lower their error rates, this implies a decrease in the respective rates will decrease revenue. From firm 1's point of view, although decreasing $FP_1$ will increase its market share, the corresponding drop in $p_1$ is larger and results in a net negative effect. 

\begin{lemma}[\text{Investments in Dominating Dimension}]
    Firm 1's revenue is strictly decreasing in FN rates, and firm 2's revenue is strictly decreasing in FP rates. This implies that investing on the strictly dominating dimension results in a strict increase in revenue. 
\end{lemma}

Similar to Lemma 1, this implies that $\frac{\partial R_1}{\partial FN_1} < 0$ and $\frac{\partial R_2}{\partial FP_2} < 0$, meaning that investments to decrease these error rates will result in revenue increases. 

\begin{theorem}[\text{Optimal Investment Choices}]
    Both firms should invest in the dimension they are strictly dominating in. In particular, firm 1 should invest to lower $FN_1$, and firm 2 should invest to lower $FP_2$. 
\end{theorem}

The theorem follows trivially from Lemmas 1 and 2. This result is slightly counter-intuitive. Firms should invest on further establishing their competitive advantage rather than trying to close the gap on the other dimension. Such investment behavior implies a degree of specialization in error rates across firms. If a firm is trying to produce a model exclusively for image classification in computer vision, it may not be very worried if the model is very inaccurate on classifying images of animals. Past research has also touched on this point. 

In addition, investing on their inferior dimension is equivalent to reducing the product differentiation in the market. This would imply that firms would no longer be able to appeal as much to extreme consumers, resulting in lower and more similar prices between firms.

\subsection{Investment Costs}

We now consider the case with an explicit investment cost function for each firm. In addition to investing on reducing their error rates explicitly, we introduce a latent variable for the size of each firm's training dataset denoted as $S$. Firms are able to invest in two ways: increasing the size of their training dataset and explicitly improving their error rates. A size investment will have more indirect benefits to the firm: any future investments on error rates will be less costly. 

Much of the artificial intelligence and machine learning literature on scaling laws (\citealp{kaplan2020scaling}; \citealp{hoffmann2022training}) has demonstrated that dataset size and performance/error are highly intertwined. In our framework, this implies that the effectiveness of firm investments are directly related to their latent dataset size variable. We propose that the training dataset size each firm has directly effects its investment costs: a firm with a much larger training dataset will be able to invest more efficiently on improving their error rates as opposed to a firm with much fewer training samples. 

Marginal costs should also be different across FP and FN rates. There should a distinct tradeoff for each firm on reducing their FP and FN rates (\citealp{tirole1988theory}; \citealp{mas1995microeconomic}). In particular, if a firm already has a low FP error rate, it should be more expensive to lower it even further. A firm with $FP = 0.1$ and investing to reduce it to 0.09 should experience much higher costs than a firm attempting to lower its FP rate from 0.3 to 0.29. 

Therefore, each firm's investment cost function has the following general form: 
$$ 
C(FP, FN, S) = \frac{g(S)}{2} \left[ s_{\overline{FP}}(FP - \overline{FP})^2 + s_{\overline{FN}}(FN - \overline{FN})^2\right] + f(S)
$$

\noindent where $f(S) = \lambda_{linear}(S - \bar{S}) + \frac{\lambda_{convex}}{p}(S - \bar{S})^p + \lambda_{runtime}(S - \bar{S})$ (with $p \geq 2$),  is an increasing convex function capturing the firm's cost for increasing their training dataset size from $\bar{S}$ to $S$, and $g(S): S \rightarrow \mathbb{R}$ is a monotonically decreasing function that maps the firm's choice of size to a cost scaling parameter. 

Here, the bar variables $\overline{FP}, \overline{FN},$ and $\overline{S}$ denote the firm's current error rates and size, respectively.\footnote{For notational clarity, we omit the firm-specific subscript $j$ for each error rate: all variables above are unique to each firm. The factor of 2 is included purely for cleanliness when differentiating.} The $\lambda$ parameters specify how sensitive the firm is to different sources of scaling costs: $\lambda_{linear}$ specifies linear scaling costs for increasing the number of training samples, $\lambda_{convex}$ specifies convex scaling costs (such as potentially regulatory or operational costs), and $\lambda_{runtime}$ specifies the additional degree of costs incurred from greater runtime during training. Finally, $s_{\overline{FP}}$ and $s_{\overline{FN}}$ capture the fact that error rates are more costly to lower when they are already low; a simple specification could set $s_{\overline{FP}} = \frac{1}{\overline{FP}}$, $s_{\overline{FN}} = \frac{1}{\overline{FN}}$. Note that $C(\overline{FP},\overline{FN},\overline S)=0$. 

We first discuss the case of split domination. Under the results from Section 4.2, neither firm will have an incentive to invest on improving the error rates on its strictly inferior dimension. Therefore, the addition of a non-zero cost function will only dis-incentivize such investments; we implicitly focus on the case of investments on one's superior dimension. 

Under this structure, firms will invest on two dimensions: their superior dimension and firm size. Although size directly affects the firm's cost for improving their error rates, it acts as a latent variable in the firm's analysis. The firm has some size level that is private information; it then chooses an optimal level of investment on both size and error rates to improve $\overline{FP}$ and $\overline{FN}$ to $FP$ and $FN$ respectively. Then, these error rates are reported to all consumers, and price equilibria are determined through market clearing. Formally, the firm's revenue is unaffected by their training dataset size. This resembles real-world scenarios in which a larger training dataset may allow for more efficient model investments, but in the end, the models themselves and their attractiveness to consumers are not directly related to the firm's training dataset size. 

We can thus compute the firm's optimal level of investment on both such dimensions using a standard optimization argument. Firms will choose investment levels such that the marginal cost and marginal revenue are equal, and we note from above that firm 1 dominates on the FN dimension and firm 2 on the FP dimension. For simplicity, we focus on the analysis from firm 1's perspective; the analysis is symmetric for firm 2. 

\begin{corollary}[\text{Optimal Choices Under AI-Based Cost Functions for Firm 1}]
    Firm 1's investment choices $(FN_1^*,S_1^*)$ are an interior optimum and  satisfy the following: 
    $$ \frac12 g'(S_1^*)\big[s_{\overline{FP_1}}(FP_1-\overline{FP_1})^2 + s_{\overline{FN_1}}(FN_1-\overline{FN_1})^2\big] + f'(S_1^*) = 0$$
    and 
$$ FN_1^* = \overline{FN_1} + \frac{\partial R_1/\partial FN_1}{g(S_1^*) s_{\overline{FN_1}}}$$
\end{corollary}

The first condition presents an optimality condition on $S_1^*$: since revenue is not directly affected by the firm's latent size variable, this is just an optimization condition on the cost function. Note that since $g(S)$ is decreasing and $f(S)$ is convex, such an $S_1^*$ exists. The second condition is a more standard firm profit maximization condition; details are provided in Appendix \ref{proof:corollary1}.

The second condition implies that the size of the firm's optimal investment is strongly dependent on its scalability and growth potential captured by $g(S)$. In particular, if a firm is able to make use of its training data very efficiently (i.e., having a lower $g(S)$), then the optimal error rate it should aim for ($FN_1^*$ in the case above) is much higher.
  
We can do a similar analysis under the strict domination case. As discussed in Proposition 2, the entrant can invest on either dimension, and it chooses the optimal investment decision depending on the ordering between the firms' FP and FN rates. Since now a lump-sum investment by the entrant on either dimension leads to successful entry, we consider investments on both dimensions. However,  unless the entrant faces a very forgiving cost function (that is, $g(S)$ is very small), entry is very difficult to incentivize. Even if the entrant does successfully obtain a non-zero market share, it will receive a very small level of revenue and is unlikely to have this offset the incurred costs. 

Building on Section 4.1, when firm 1 strictly dominates firm 2, a profitable investment tuple $(FP_2,FN_2,S_2)$ by firm 2 must change the ordering of error rates sufficiently to produce interior positive prices, as in Proposition 2. In addition, the resulting post-investment revenue should be greater than the incremental investment cost. 

\section{Welfare Analysis}

We have shown the existence of strictly profitable investments for firms, and we now analyze the welfare implications of such investments. When firms invest on their superior dimension, they increase the level of differentiation, and so both firms charge higher prices.

\begin{claim}
    Strictly profitable firm investments have a negative impact on consumer welfare and a strictly positive effect on firm welfare. The overall effect on welfare is positive (i.e., firm welfare + consumer welfare $>$ 0).
\end{claim}

In particular, such an investment not only increases revenue for the investing firm but also for the other firm. This is due to the fact that although the other firm loses a portion of its market share, it is also able to increase prices. 

Consumers, though, are adversely affected by such profitable investments due to the increases in price. If firm 1 invests on lowering $FN_1$ (i.e., a profitable investment), both firms' prices increase and the indifferent consumer shifts to the right, meaning firm 1 captures more market share. However, total consumer costs increase as well. Extreme consumers (consumers with $\alpha$ close to 0 or 1) purchase from the same firm and are strictly worse off due to higher prices. Since firm 1 charges a lower price, the consumers that move from firm 2 to firm 1 will be better off, but this proportion of moderate consumers is relatively small and does not offset the negative effects on other consumers. 

In particular, overall consumer welfare is measured by total costs across all consumers: 
\[
\text{Welfare} = \int_0^{\tilde{\alpha}} c_i(1) \, d\alpha + \int_{\tilde{\alpha}}^1 c_i(2) \, d\alpha,
\]
where
\[
c_i(1) = \alpha \Delta F_1 + p_1^* + FN_1, \quad c_i(2) = \alpha \Delta F_2 + p_2^* + FN_2.
\]

\begin{claim}
    Firm investments on their inferior dimension have a strictly positive effect on consumers but a negative effect on overall welfare. 
\end{claim}

For example, an investment by firm 1 towards lowering its FP rate (its inferior dimension) would result in a strictly lower profit as shown by Theorem 4. Furthermore, firm 2 would experience a relatively larger drop in profit, due to losing market share but also having to reduce its price. However, consumers would be strictly better off due to lower prices. An interesting consequence of this is that firms can be incentivized to invest on their inferior dimension, which would be equivalent to reducing product differentiation in the market. 

\section{Summary and Conclusions}

This paper develops a consumer-firm duopoly model to analyze optimal firm behavior through specialized investments. Each firm aims to minimize its model's error, but this choice can often be suboptimal when firms compete and consumers have heterogeneous preferences. By decomposing model error into false positive and false negative rates, we allow firms to reduce error on one dimension at a time through investments. 

Our key finding is that when firms face a trade-off in how they allocate their efforts to reduce errors in different dimensions (e.g., FP versus FN), the optimal strategy is not to close performance gaps evenly but rather to invest further in the dimension where they already hold a clear advantage. This approach leads to greater product differentiation, higher equilibrium prices, and ultimately higher firm revenues \textemdash albeit at the cost of lower consumer welfare. These findings directly relate back to the motivations laid out in the introduction, highlighting the practical difficulty and subjectivity in minimizing model errors. Rather than striving to minimize all errors at once, firms benefit from specialized improvements that align with their consumers' cost sensitivities and their own competitive strengths.

Our model assumes all information about consumer types is public information; introducing an element of private information into the consumer types could be interesting. Each consumer would be endowed with their own personal FP/FN rates and only buy a product from firms if the transaction is profitable error-wise. We assume that every consumer must not only purchase from a firm, but we can also deterministically pinpoint from which firm they will buy from. Implicitly, we assume a mature and entirely covered market in which we not only have a fixed market size with no growth but also enforce that all firms in the market are actually selling a product. A reasonable relaxation of the model could consider the possibility of shut-downs when their product is no longer profitable, and similarly the potential for market entry if a firm comes up with an innovative product. 

\clearpage
\bibliography{bibliography.bib}
\appendix

\clearpage
\section{Derivations}
\subsection{Equilibrium Prices in Perfectly Positively Correlated Model: $\alpha = \beta$}

Equilibrium prices are derived in the canonical way: determine the indifferent consumer, solve for each firm's reaction function, and compute resulting equilibrium prices. 
To determine the indifferent consumer $\Tilde{\alpha}$, we find the corresponding value on the unit interval such that that consumer is indifferent between buying from firm 1 and firm 2: $c_i(1) = c_i(2)$. We now consider three cases on $\mathcal{F}_1 - \mathcal{F}_2$: (i) $\mathcal{F}_1 - \mathcal{F}_2 > 0$, (ii) $\mathcal{F}_1 - \mathcal{F}_2 < 0$, and (iii) $\mathcal{F}_1 - \mathcal{F}_2 = 0$. Note that case (i) and (ii) are almost identical due to symmetry. We present the derivation for case (i), case (ii) is symmetric. Case (iii) is straightforward: if both firms have same total error, the resulting equilibria will be both charging the highest possible price $\Bar{p}$. 

$\Rightarrow {\Tilde{\alpha}} \mathcal{F}_1 + p_1 = {\Tilde{\alpha}} \mathcal{F}_2 + p_2  \Rightarrow \Tilde{\alpha} = \frac{p_2 - p_1}{\mathcal{F}_1 - \mathcal{F}_2}$. This expression is then bounded to the unit interval, taking on boundary values 0 and 1 otherwise:
\[
    \Tilde{\alpha} =\begin{cases} 
    0 & \text{if}  \ \frac{p_2 - p_1}{\mathcal{F}_1 - \mathcal{F}_2} \leq 0 \\
    \frac{p_2 - p_1}{\mathcal{F}_1 - \mathcal{F}_2} & \text{if}  \ 0 < \frac{p_2 - p_1}{\mathcal{F}_1 - \mathcal{F}_2} < 1 \\
    1 & \text{if} \ \frac{p_2 - p_1}{\mathcal{F}_1 - \mathcal{F}_2} \geq 1.  \\
   \end{cases}
\]

Any consumers with $\alpha < \Tilde{\alpha}$ will buy from firm 1, and all others buy from firm 2. 

We then determine firm 1 and firm 2's total revenue as: 
\begin{equation}
\begin{aligned}
        R_1(p_1, p_2) &= \Tilde{\alpha}p_1 = \frac{p_2 - p_1}{\mathcal{F}_1 - \mathcal{F}_2} \times p_1,
        \\R_2(p_1, p_2) &= (1-\Tilde{\alpha})p_2 = (1 - \frac{p_2 - p_1}{\mathcal{F}_1 - \mathcal{F}_2}) \times p_2. \nonumber
\end{aligned}
\end{equation}

Computing and solving the FOCs for each firm yields the reaction functions $p_1^*(p_2) = \frac{1}{2}p_2$ and $p_2^*(p_1) = \frac{\mathcal{F}_1 - \mathcal{F}_2 + p_1}{2}$, and solving the two equations yields the following equilibrium prices:
\begin{equation}
\begin{aligned}
        p_1 = \frac{1}{3}(\mathcal{F}_1 - \mathcal{F}_2), \\ 
        p_2 = \frac{2}{3}(\mathcal{F}_1 - \mathcal{F}_2). \nonumber
\end{aligned}
\end{equation}
An identical derivation can be shown for the case where $\mathcal{F}_1 - \mathcal{F}_2 < 0$, and prices will simply be reversed:
\begin{equation}
\begin{aligned}
        p_1 = \frac{2}{3}(\mathcal{F}_2 - \mathcal{F}_1), \\ 
        p_2 = \frac{1}{3}(\mathcal{F}_2 - \mathcal{F}_1). \nonumber
\end{aligned}
\end{equation}

\subsection{Equilibrium Prices with Split Domination in Perfectly Negatively Correlated Model: $\alpha = 1 -  \beta$}

We assume the following ordering on FP and FN rates: $FN_1 <  FN_2, FP_1 > FP_2$. Under this assumption, note that $\Delta F_1 - \Delta F_2$ is guaranteed to be non-negative by definition: $\Delta F_1 - \Delta F_2 = (FP_1 - FP_2) + (FN_2 - FN_1) > 0$. We then can solve for the indifferent consumer by equating cost functions: 

$\Rightarrow {\Tilde{\alpha}} \Delta F_1 + p_1 + FN_1 = {\Tilde{\alpha}}  \Delta F_2 + p_2 + FN_2 \Rightarrow \Tilde{\alpha} = \frac{p_2 - p_1 + FN_2 -  FN_1}{\Delta F_1 - \Delta F_2}$.

However, note that this expression for the indifferent consumer $\Tilde{\alpha}$ must be bounded by 0 and 1 inclusive. If $\Tilde{\alpha} < 0$, no consumers are buying from firm 1 and the indifferent consumer is effectively at 0. Analogously, if $\Tilde{\alpha} > 1$ implies all consumers would buy from firm 1, and thus the indifferent consumer is at 1. $\Tilde{\alpha}$ can thus be written as: 
\begin{equation}
     \Tilde{\alpha} =\begin{cases} 
      0 & p_2 - p_1 \leq FN_1 - FN_2 \\
      1 & p_2 - p_1 \geq FP_1 - FP_2 \\
      \frac{p_2 - p_1 + FN_2 - FN_1}{\Delta F_1 - \Delta F_2} & \text{otherwise.}   \nonumber \\
   \end{cases}
\end{equation}

Firm revenue can be expressed as proportion of buying consumers times price, where revenue is increasing in $\Tilde{\alpha}$ for firm 1, and similarly decreasing in $\Tilde{\alpha}$ for firm 2. 
\begin{equation}
\begin{aligned}
        R_1(p_1, p_2) &= \Tilde{\alpha}p_1 \nonumber
        \\R_2(p_1, p_2) &=  (1-\Tilde{\alpha})p_2
\end{aligned}
\end{equation}

The firms' reaction functions can then be solved using casework on $\Tilde{\alpha}$ and relevant FOCs. Note that if $\Tilde{\alpha} = 0$, then $R_1 = 0, R_2 = p_2$. Since firms are not subject to direct costs in this model, firm 2 will set price $p_2 = \Bar{p}$ and $p_1 = 0$. Similarly, if $\Tilde{\alpha} = 1$, firm 2's revenue will be 0 and equilibrium prices will be $p_1 = \Bar{p}$ and $p_2 = 0$. When $0 < \Tilde{\alpha} < 1$, the reaction functions can be calculated by taking the FOCs of firm revenues with respect to prices.
\begin{equation}
    \begin{aligned}
    p_1^*(p_2) = \max \{0, \frac{FN_2 - FN_1 + p_2}{2}\}; \\
    p_2^*(p_1) = \max\{0, \frac{FP_1 - FP_2 + p_1}{2}\}. \nonumber
    \end{aligned}
\end{equation}

To solve for equilibrium prices, we solve the two firms reaction functions as a system of equations, which yields:
\begin{equation}
\begin{aligned}
        p_1^* &= \frac{2(FN_2 - FN_1) + FP_1 - FP_2}{3}; \\
        p_2^* &= \frac{FN_2 - FN_1 + 2(FP_1 - FP_2)}{3} .\nonumber
\end{aligned}
\end{equation}

\subsection{Equilibrium Prices with Strict Domination in Perfectly Negatively Correlated Model: $\alpha = 1 -  \beta$}

The general framework is very similar to A.2, but the conditions of $\Tilde{\alpha}$ are slightly different due to different orderings on FP and FN rates. The indifferent consumer $\Tilde{\alpha}$ is solved as in A.2: 
\begin{align*}
    \Tilde{\alpha} = \frac{p_2 - p_1 + FN_2 - FN_1}{\Delta F_1 - \Delta F_2}.
\end{align*}

Note that since $|FN_2 - FN_1| < |FP_2 - FP_1|$, the denominator of this expression is negative. Therefore, to result in a non-negative $\Tilde{\alpha}$, we must have that $p_2 - p_1 + FN_2 - FN_1 \leq 0 \Rightarrow p_1 \geq p_2 + FN_2 - FN_1$, meaning that $p_1$ would be strictly larger than $p_2$ as $FN_2 - FN_1$ is positive by assumption. This makes intuitive sense as well; since firm 1 has a better product, they charge a higher price to consumers while also maintaining a significant consumer share. We also want to upper bound $\Tilde{\alpha}$ by 1, and solving gives that $p_1 \leq p_2 + FP_2 - FP_1$, which limits viable non-negative prices to an interval. Note that max operators are implicitly present to ensure non-negativity in both prices; they are omitted for conciseness. This condition ensures that the gap in between $p_1$ and $p_2$ is not too large. 
Thus, $\Tilde{\alpha}$ can be written as: 
\begin{equation}
     \Tilde{\alpha} =\begin{cases} 
      0 & p_2 - p_1 > FN_1 - FN_2 \\
      1 & p_2 - p_1 < FP_1 - FP_2 \\
      \frac{p_2 - p_1 + FN_2 - FN_1}{\Delta F_1 - \Delta F_2} & \text{otherwise.}  \\
   \end{cases}
\end{equation}

The reaction functions are similar to A.2; the main difference comes from solving for the resulting equilibrium. Since $FP_1 - FP_2 < 0$, firm 2's reaction function has a negative intercept, and so firm 2 chooses $p_2 = 0$ if firm 1's price is small enough.  

\section{Proofs}
\subsection{Theorem 1}
Since FP and FN rates are now adjustable, the reaction functions are modified as follows: 
\begin{equation}
    \begin{aligned}
    p_1^*(p_2, \mathcal{P}, \mathcal{N}) = \max \{0, \frac{FN_2 - FN_1 + p_2}{2}\}, \\
    p_2^*(p_1, \mathcal{P}, \mathcal{N}) = \max\{0, \frac{FP_1 - FP_2 + p_1}{2}\},
    \end{aligned}
\end{equation}
where $\mathcal{P}, \mathcal{N}$ represent the set of FP and FN rates respectively. In the two firm case, $\mathcal{P} = \{FP_1, FP_2\}$ and $\mathcal{N} = \{FN_1, FN_2\}$. Rewrite the traditional reaction functions presented the split domination case using representative constants: 
\begin{equation}
    \begin{aligned}
        p_1^*(p_2, \mathcal{P}, \mathcal{N}) = \max \{0, \frac{a + p_2}{2}\}, \\
        p_2^*(p_1, \mathcal{P}, \mathcal{N}) = \max\{0, \frac{b + p_1}{2}\}, \nonumber
    \end{aligned}
\end{equation}
where $a = FN_2 - FN_1 > 0, b = FP_1 - FP_2 < 0$. To ensure non-zero profits for firm 2, $p_2^*$ must be non-zero. Thus, for calculation the max operators can be ignored, and we can solve for $p_2^*$ in terms of $a,b$:
\begin{align*}
    \frac{b+p_1}{2} &= p_2 \rightarrow p_1 = 2p_2 - b \\
    &\Rightarrow 2p_2 - b = \frac{a + p_2}{2} \\
    &\Rightarrow p_2^* = \frac{a + 2b}{3}
\end{align*}

For this expression to be positive, we must have that $a + 2b > 0$. Note that $a > 0, b < 0$, and since firm 2 cannot increase a (increasing error rate is not beneficial), their only choice is to decrease b sufficiently. In particular, we must have that $b > -\frac{a}{2} \Rightarrow b < \frac{a}{2}$. This is by definition our desired claim: if firm 2 can reduce the gap in FP rates sufficiently so that $FP_1 - FP_2 = b$ is less than $\frac{a}{2} = \frac{FN_2 - FN_1}{2}$, $p_2^*$ ends up being strictly positive and results in positive profits. 

Following the choices of FP and FN rates from Figure 3, such a shift makes firm 2 strictly better off and firm 1 strictly worse off. Changes in resulting revenue for firm 1 and firm 2 are presented below:
\begin{align*}
   & \Bar{R_1}(p, \mathcal{P}, \mathcal{N}) = \left(\frac{p_2^* - p_1^* + FN_2 - FN_1}{\Delta F_1 - \Delta F_2}\right)p_1^* = 0.833 \times 0.167 = 0.1388 < 0.15; \\
   & \Bar{R_2}(p, \mathcal{P}, \mathcal{N}) = \left(1 - \frac{p_2^* - p_1^* + FN_2 - FN_1}{\Delta F_1 - \Delta F_2}\right)p_2^* = 0.167 \times 0.033 = 0.0055 > 0.
\end{align*}

\subsection{Theorem 2}

To prove the Optimal Investment Choices theorem, we simply prove Lemma 1 and Lemma 2, and the theorem follows trivially. We differentiate the revenue expression with respect to the corresponding error rate and arrive at our desired claim with sign analysis. 

We focus on Lemma 1; similar sign analysis can be conducted with respect to the other error rate for Lemma 2. Consider $\frac{\partial R_1}{\partial FP_1}$. We simply substitute in our expressions for the equilibrium price and indifferent consumer and differentiate: 
\begin{align*}
    &R_1 = \tilde{\alpha}p_1 = \left( \frac{p_2-p_1 + FN_2 - FN_1}{\Delta F_1 - \Delta F_2} \right) \left(\frac{2(FN_2 - FN_1) + FP_1 - FP_2}{3} \right) \\
    &\Rightarrow \frac{\partial R_1}{\partial FP_1} = \frac{\partial}{\partial FP_1} (\tilde{\alpha}) * p_1 + \frac{\partial}{\partial FP_1} (p_1) * \tilde{\alpha} \\
    &= \left( \frac{FN_2 - FN_1}{(\Delta F_1 - \Delta F_2)^2} \right)\left(\frac{2(FN_2 - FN_1) + FP_1 - FP_2}{3} \right) + \frac{1}{3} \left ( \frac{FP_1 - FP_2}{\Delta F_1 - \Delta F_2} \right)
\end{align*}
where the simplifications for the indifferent consumer term follow when plugging in our expression for prices and simplifying terms. If we use our assumptions that $FN_1 < FN_2, FP_1 > FP_2$ and $|FN_2 - FN_1 | < |FP_2 - FP_1|$, then we can see that this expression is always positive. In particular, prices are always non-negative, and the numerators in both the other terms are positive, so the overall derivative is positive. However, since we invest to reduce our error rate, a reduction in $FP_1$ for firm 1 will result in strictly lower revenue. This shows that for firm 1, investing on improving their inferior dimension, i.e. false positive rates, will result in strictly lower profits. A similar analysis can be conducted to show $\frac{\partial R_2}{\partial FN_2} > 0$, meaning firm 2 also gets strictly lower revenue when investing on their dominated dimension. The only difference in firm 2's case is substituting in a $1 - \tilde{\alpha}$ term and $p_2$ instead of $p_1$. 

If we repeat the same analysis for each firm's superior dimension, then we see that both firms benefit from investing: 
$$\frac{\partial R_1}{\partial FN_1} < 0, \frac{\partial R_2}{\partial FP_2} < 0.$$

\subsection{Corollary 1}
\label{proof:corollary1}

We first explicitly derive an expression for $\frac{\partial R_1}{\partial FN_1}$ using a similar stream of logic to Theorem 2. In particular, let $a = FN_2 - FN_1 > 0, b = FP_1 - FP_2 < 0$ as in the proof for Theorem 1. Thus, from Claim 2 we have that prices are $p_1  = \frac{2a + b}{3}$ and $ p_2 = \frac{a + 2b}{3}$ and the general expression for revenue $R_1$ as:

$$R_1 = \tilde{\alpha}p_1 = \left( \frac{p_2-p_1 + a}{\Delta F_1 - \Delta F_2} \right) \left (\frac{2a + b}{3} \right)$$

We then want to compute $\frac{\partial R_1}{\partial FN_1}$:

\begin{align*}
\frac{\partial R_1}{\partial FN_1} =& \frac{\partial}{\partial FN_1} (\tilde{\alpha}) * p_1 + \frac{\partial}{\partial FN_1} (p_1) * \tilde{\alpha} \\ \\
=& \left(-\frac{b}{3(\Delta F_1 - \Delta F_2)^2} \right)\left (\frac{2a + b}{3} \right) - \frac23 \left( \frac{2a + b}{3(\Delta F_1 - \Delta F_2)}\right) \\ \\ 
= & \frac{(-1) \cdot (2a- b) \left[b + 2(\Delta F_1 - \Delta F_2)\right]}{3(\Delta F_1 - \Delta F_2)^2}
\end{align*}
which is always negative as expected. Then, consider the standard firm profit maximization problem and the corresponding FOC:
$$
\Pi_1 = R_1(p_1, p_2) - C_1(FP_1, FN_1, S_1)
$$
\begin{align*}
    \frac{\partial \Pi_1}{\partial FN_1} =& \frac{\partial R_1}{\partial FN_1} - \frac{\partial C_1}{\partial FN_1} = 0 \\ \\ 
    \Rightarrow& \frac{(-1) \cdot (2a- b) \left[b + 2(\Delta F_1 - \Delta F_2)\right]}{3(\Delta F_1 - \Delta F_2)^2} = g(S_1) s_{\overline{FN_1}} (FN_1 - \overline{FN_1})
\end{align*}

which gives the desired relation once we substitute back in the expressions for $a$ and $b$. The analysis for the optimal size $S_1$ is straightforward: since the size variable is latent and does not directly affect firm revenue, we simply differentiate the cost function with respect to $S_1$ and set it equal to zero. 

\section{Additional Graphs}
\label{appendix:graphs}

We present a visual interpretation of Claim 4 obtained through numerical simulation. We construct a grid space on $\zeta$ with increments of 0.01 and plot the corresponding price and revenue under equilibria. Note that both price and revenue are highest when consumers all have negatively correlated cost sensitivities. 

\begin{figure}[!htbp]
    \centering
    \includegraphics[width=1\linewidth]{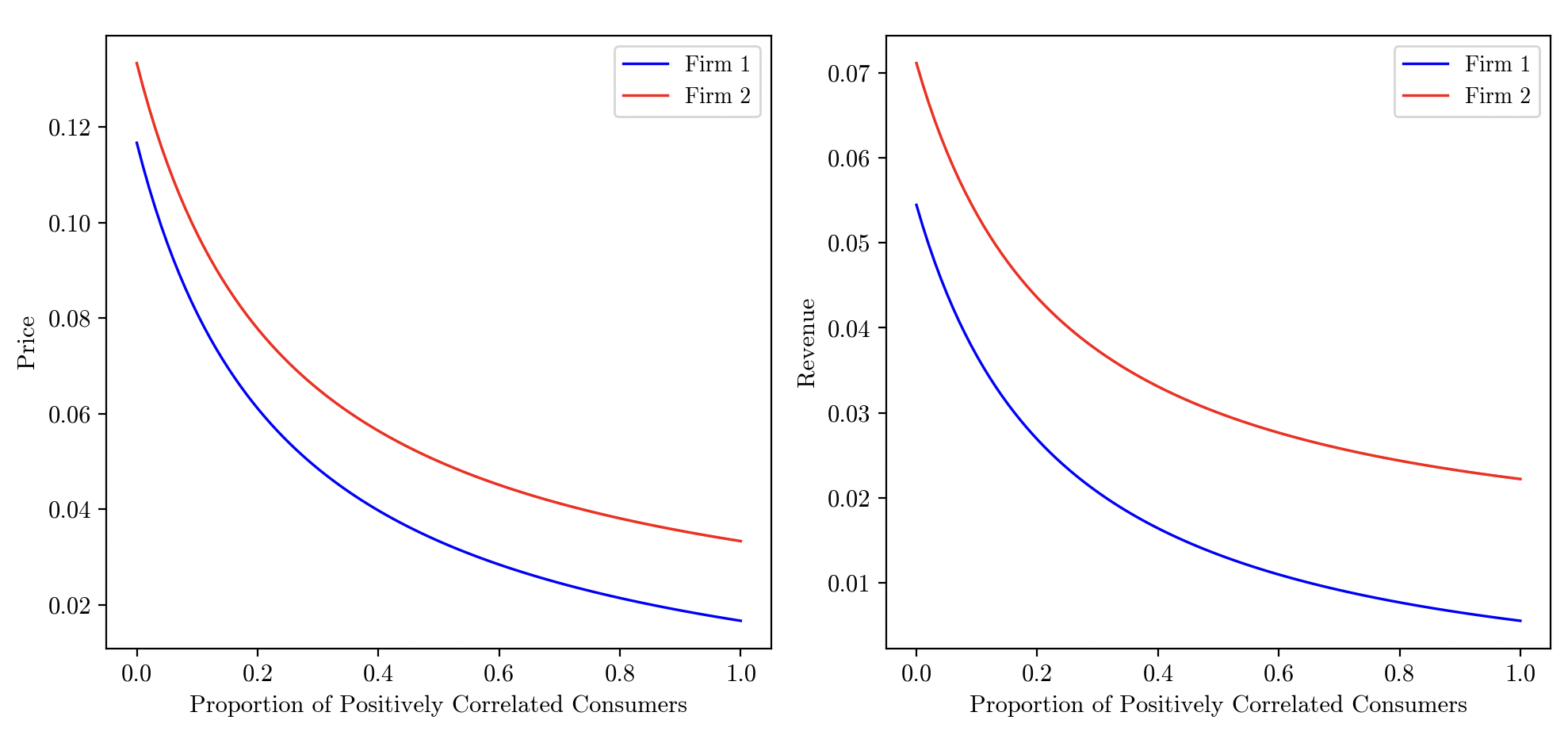}
    \caption{Price and revenue of both firms across varying proportions of positively correlated consumers $\zeta$}
    \label{fig:zeta_price}
\end{figure}

\end{document}